\documentclass[prb,twocolumn,superscriptaddress,showpacs,floatfix]{revtex4}
\usepackage[english]{babel}
\usepackage[dvips]{graphicx}
\usepackage{amsmath}
\usepackage{amsfonts}
\usepackage{amssymb}
\usepackage{placeins}

\begin{document}
\title{Selective excitations of transverse vibrational modes of a carbon 
nanotube through a ``shuttle-like'' electromechanical instability}
\author{F. Santandrea \\ \textit{Department of Physics, University of Gothenburg, SE - 412 96 
G{\"o}teborg, Sweden}} 
\noaffiliation
%\affiliation{Department of Physics, University of Gothenburg, SE - 412 96 
%G{\"o}teborg, Sweden}

\begin{abstract}
We study the dynamics of transverse oscillations of a suspended carbon 
nanotube into which current is injected from the tip of a scanning tunneling 
microscope (STM). In this case the correlations between the displacement of 
the nanotube and its charge state, determined by the position-dependent 
electron tunneling rate, can lead to a ``shuttle-like'' instability 
for the transverse vibrational modes. We find that selective excitation of a 
specific mode can be achieved by an accurate positioning of the STM tip. 
This result suggests a feasible way to control the dynamics of this 
nano-electromechanical system (NEMS) based on the ``shuttle instability''.
\end{abstract}


\maketitle

There are several reasons for the considerable current interest in 
nano-electromechanical systems (NEMS), both for technological applications 
and fundamental research. The peculiar combination 
of several features such as high vibrational frequencies and 
small masses which characterize most NEMS makes these systems 
very suitable for the realization of new measurement tools with 
extremely high sensitivity in mass sensing and force 
microscopy applications [\onlinecite{Roukes2000, Blencowe2005}]. Furthermore, the 
mechanical elements of the NEMS (typically cantilevers or beams) 
are considered the most promising structures where quantum features of 
motion could be experimentally detected [\onlinecite{Schwab2005}].

The physical basis for many of the interesting functionalities of NEMS is the 
strong interplay between mechanical and electronic degrees of freedom 
[\onlinecite{Poncharal1999,Purcell2002,Sapmaz2003,Sazonova2004}]. 
In the particular case of a nano-electromechanical single-electron 
transistor device having a metallic dot as movable part, the equilibrium
position of the dot can become unstable as a consequence of the 
electromechanical coupling. In this case the dominant mechanism 
for the transport of charge is based on the oscillations of the dot 
which can ``shuttle'' the tunneling electrons across the system  
[\onlinecite{Gorelik1998, Moskalenko2009}].

The typical set-up for many NEMS includes a spatially extended movable element 
such as a suspended carbon nanotube, whose dynamics has been demonstrated to be
characterized by a number of different vibrational modes [\onlinecite{Huttel2008}]. 
The relevance of many mechanical modes in the transport of charge suggests 
that the variety of effects due to the electromechanical coupling in 
suspended carbon nanotube-based NEMS may be even richer than in the 
ordinary ``shuttle'' system. Jonsson \emph{et al.} have shown 
[\onlinecite{Jonsson2005,Jonsson2007,Jonsson2008}] that if extra charge 
is injected into the movable part of the device from the tip of a 
scanning tunneling microscope (STM) a nano-electromechanical 
``shuttle-like'' instability can be induced for the transverse 
vibrational modes of the nanotube. 

The selective promotion of the electromechanical instability for different
vibrational modes provides an interesting perspective for probing the dynamics 
of NEMS. Here we show that such selective excitation can be achieved by means 
of local injection of electric charge. The main idea is to optimize the 
electromechanical coupling for the mode(s) which we want to make unstable. 
The local character of the electric charge injection makes the selective 
excitation of the nanotube transverse modes possible by varying the 
position of the STM tip.

\begin{figure}
\center
\includegraphics[width=0.5\textwidth]{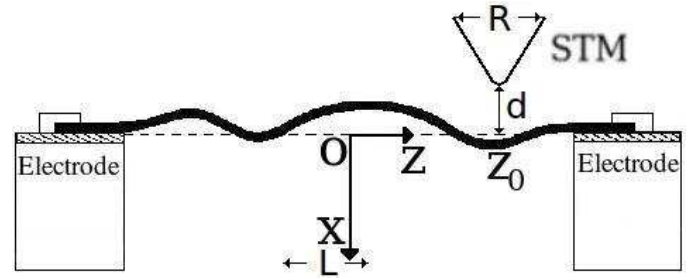}
\caption{Sketch of the model system considered. The distance of the suspended 
carbon nanotube from the STM tip affects the electron tunneling rate between 
them, while the tunneling rate between the nanotube and the leads is constant
%The right lead can be used as a gate electrode in order to control the number 
%of extra electrons tunneling on the nanotube like in an ordinary single-electron
%transistor setup 
(the vibration amplitude and the STM-nanotube equilibrium distance $d$ shown are exaggerated
for clarity).}
\label{fig:system}
\end{figure}

We will consider here the same device analyzed by Jonsson \emph{et al.} 
since it provides a convenient set-up to control the electromechanical 
coupling of different vibrational modes. The system is sketched in 
Fig. (\ref{fig:system}) and it consists of a suspended metallic carbon 
nanotube in tunneling contact with an STM tip and one supporting lead. 
%A bias voltage $V$ is applied between the tip and the lead so 
%that electrons can tunnel across the nanotube, while another electrode 
%with potential $V_g$ (indicated as 'Gate' in Fig. (\ref{fig:system})) 
%is capacitively coupled to the nanotube. 

We take the $z$-axis along the nanotube axis, while its cross section 
lies in the $xy$ plane. The STM tip is put over point (0, 0, $z_0$) 
and its distance from the nanotube at equilibrium is $d \simeq $1 nm. 

In order to describe the motion of the nanotube we model it as a classical 
elastic beam of length $L$ clamped at both ends and focus on its flexural 
vibrations.
%This kind of approach has provided results that are quite consistent with 
%experimental data in all the existing works about carbon nanotube-based 
%NEMS (see for example [\onlinecite{Poot2007}] and references therein). 
%Throughout this work we neglect the deformation of the cross-section 

The motion of the nanotube in the $xz$ plane can be described through 
$u(z,t)$, its displacement along the $x$ axis from the static equilibrium 
configuration. If the amplitude 
of the oscillations is small enough for linear elasticity theory to be valid, 
the time evolution of $u(z,t)$ is determined by the following equation 
[\onlinecite{Cleland2003}]:

\begin{equation} \label{eq:elast}
\rho S \frac{\partial^2 u}{\partial t^2}+\frac{EI}{L^4}
\frac{\partial^4 u}{\partial z^4}=F_x^{\rm el}(N(t),z,z_0).
\end{equation}

\noindent In Eq. (\ref{eq:elast}), $\rho$ is the carbon nanotube density, 
$S$ the cross section, $E$ the Young modulus, $I$ the cross section moment of inertia, $N(t)$ the number 
of extra electrons on the nanotube at time $t$ and $F_x^{\rm el}$ the $x$ 
component of the external force (per unit length) generated by the electrostatic 
interaction between the STM tip and the nanotube. 

The precise spatial distribution of $F_x^{\rm el}$ depends on the details of 
the geometric structure of the tip apex. However, a simple electrostatic 
analysis indicates that for $|z-z_0| \gg R$, $d$ (where $R \lesssim 10^{-8}$ m 
is the effective size of the STM tip), the force $F_x^{\rm el}$ decays at least
as $|z-z_0|^{-3}$. Therefore the influence of the metallic leads can be ignored
as long as the STM tip is not too close to them and we can write 
$F_x^{\rm el}(N,z,z_0) \simeq F_x^{\rm el}(N,z-z_0)$.

The displacement field $u(z,t)$ and the force per unit length $F_x^{\rm el}$ 
can be expressed as linear combinations of eigenfunctions $\varphi_j(z)$ of the
operator $d^4/dz^4$ with the boundary conditions $u(0,t)=u(L,t)=0$, 
$du/dz(0,t)=du/dz(L,t)=0$.

The expansion of $u(z,t)$ and $F_x^{\rm el}$ over the complete set of functions
$\varphi_j(z)$ makes it possible to decompose Eq. (\ref{eq:elast}) into a set of 
equations of motion for the eigenmode amplitudes $x_j(t)$, which can be written 
in hamiltonian form by introducing the conjugate momenta $\pi_j(t)$: 

\begin{subequations} \label{eq:mode_expansion}
\begin{align}
\dot{x}_j & = \frac{\pi_j}{m} \\
\dot{\pi}_j & + \gamma_j \pi_j + m\omega_j^2 x_j = Lf_j^{\rm el}(N(t),\{x_j\},z_0). 
\label{eq:x_j}
\end{align}
\end{subequations}

\noindent In Eqs. (\ref{eq:mode_expansion}), $f_j^{\rm el}$ are the coefficients in the
expansion of $F_x^{\rm el}$ over the complete set of functions $\varphi_j(z)$
(which are chosen to be dimensionless), $m=\rho S L$ is the mass of the nanotube 
and $\omega_j$ are the frequencies of
the transverse vibrational modes, given by 
$\omega_j=\sqrt{k_jEI/(\rho S L^2)}$, where the eigenvalues $k_j$ can be found 
by solving: $\textrm{cos}k_j^{1/4} \textrm{cosh}k_j^{1/4}=1$.

We introduced in Eq. (\ref{eq:x_j}) a phenomenological damping force for each mode, 
$-\gamma_j\pi_j$, where $\gamma_j$ has the dimension of 
inverse time. The motion of the nanotube 
is inevitably affected by dissipative mechanisms, which can be 
related to its coupling to the environment and to several internal processes. 
%It is in general quite hard to identify the 
%dominant dissipative mechanism in a given NEMS. 
We will later consider a general form for $\gamma_j$ which can be used to 
describe the damping induced by a wide class of phenomena. 

An approximate expression for the force coefficients $f_j^{\rm el}$ in Eq. 
(\ref{eq:x_j}) can be found through some physical considerations on the 
characteristic lengths of the system. Since the eigenfunctions $\varphi_j(z)$ 
vary appreciably only over distances of the order of $L$, we can express each 
$f_j^{\rm el}$ as a sum of a sharply localized contribution at $z_0$ plus a 
correction: 

\begin{align} \label{def:el_force}
Lf_j^{\rm el} & =\int_{-L/2}^{L/2} F_x^{el}(N(t),z'-z_0)\varphi_j(z')dz' = 
\nonumber \\
{} & = eN(t)\mathcal{E}\varphi_j(z_0) + O \left( R^2/L^2 \right),
\end{align}

\noindent where $\mathcal{E}$ is a phenomenological parameter
which provides the magnitude of the effective electrostatic 
field between the STM and the nanotube.

The size of the correction in Eq. (\ref{def:el_force}) can be estimated in terms of the 
characteristic lengths of the system: for the typical values $L \sim 10^{-7}$ m
and $R \sim 10^{-8}$ m, the condition $R^2/L^2 \ll 1$ is fulfilled and that
defines the range of validity of the approximation $Lf_j^{\rm el} \simeq eN(t)
\mathcal{E}\varphi_j(z_0)$ which we will use from now on. 

%For what concerns the transport of charge the system is equivalent to a single
%electron transistor (SET) [\onlinecite{Ferry1997}], having one tunnel junction 
%between the STM tip and the nanotube, one tunnel junction between the nanotube 
%and the lead and a gate electrode through which it is possible to control the number
%of extra electrons that can tunnel into the nanotube. 

For what concerns the transport of charge the system is equivalent to a double
tunnel junction, having one potential barrier localized between the 
STM tip and the nanotube and the another one between the nanotube 
and the leads. In our analysis we will consider the case of electrons for 
which the inverse characteristic time of decoherence 
is much shorter than the tunneling rates, so that the description of tunneling 
as a stochastic (rather than coherent) process is sufficient.
 
Following the approach presented in [\onlinecite{Armour2004}] for a point-like 
oscillator coupled to a single-electron transistor, we define a probability density in the 
phase space of the system $P_N(\vec{x},\vec{\pi},t)$ such that $P_N(\vec{x},\vec{\pi},t)
dx_1d\pi_1dx_2d\pi_2\ldots$ (with all the $x_j$ and $\pi_j$ scaled by proper
dimensional factors) is the joint probability that at time $t$ there are
$N$ electrons in excess on the nanotube while the eigenmode amplitudes and 
momenta $\{x_j\}$, $\{\pi_j\}$ take values in the phase space region defined by
$\prod_j dx_j d\pi_j \equiv d\vec{x}d\vec{\pi}$.

We consider the system in the Coulomb blockade regime %[\onlinecite{Ferry1997}] 
and limit to one the maximum number of extra electrons on the nanotube, therefore
only the probability densities $P_0(\vec{x},\vec{\pi},t)$ and 
$P_1(\vec{x},\vec{\pi},t)$ play a role in the description of the nanotube 
dynamics. 

%In order to have $N$ =0,1 the values of the bias voltage
%and the temperature $T$ must fall into the range: 
%$E_{\rm C}<k_{\rm B}T,eV< 2E_{\rm C}$, where the charging 
%energy $E_{\rm C}$ is of the order of 0.1 eV [\onlinecite{LeRoy2004a}].

%Mention gate voltage?
%if the nanotube length is of the order of $10^{-7}$ m [\onlinecite{LeRoy2004a}].  

%The regime of single-electron charging of the nanotube can be
%attained if the bias voltage and the temperature are such that 
%$eV$, $k_{\rm B}T \ll \Delta$, where $\Delta$ is the energy spacing between the
%electronic states of the nanotube, which turns out to be of the order of a few 
%meV for nanotubes whose length is of the order of $10^{-7}$ m [\onlinecite{Sapmaz2005}]. 
%In this regime only a single electronic state is involved in the 
%transport of charge and the tunneling rates across the junctions 
%do not depend on the bias voltage. 
The coupling between the mechanical and electronic degrees of freedom arises 
because the tunneling rate between the STM tip and the nanotube, 
$\Gamma_{\rm STM-NT}(\vec{x},z_0)$ is affected by their relative distance at point
$z_0$: $\Gamma_{\rm STM-CNT}(\vec{x},z_0) = \Gamma_0 
\exp((-d+\sum_j x_j\varphi_j(z_0))/ \lambda)$, where $\lambda$ is the effective
tunneling length of the STM-nanotube junction and $\Gamma_0$ is a constant.
The factor $\Gamma_0^*(d) \equiv
\Gamma_0\exp{(-d/\lambda)}$ is the tunneling rate that would characterize the 
junction if the motion of the nanotube could be neglected. The tunneling rate 
between the nanotube and the leads, $\Gamma_{\rm NT-L}$ does not depend on the
nanotube displacement. 

We remark that all the tunneling rates are generally functions of the bias voltage. 
However, since here we always assume $V$ fixed at some value 
we never explicitly indicate the dependence on $V$. In order to be 
consistent with the condition of single-electron charging of the 
nanotube, the electron temperature and the bias voltage have to 
be in the range: $k_{\rm B}T < E_{\rm C} < eV < 2E_{\rm C}$,
where $E_{\rm C}$ is the energy required to add one extra electron 
to the nanotube. 
  
The time evolution of the probability densities $P_+(\vec{x},\vec{\pi},t) 
\equiv P_1(\vec{x}, \vec{\pi},t) + P_0(\vec{x},\vec{\pi},t)$ and $P_-(\vec{x},
\vec{\pi},t) \equiv P_1(\vec{x}, \vec{\pi},t) - P_0(\vec{x},\vec{\pi},t)$ is 
determined by the equations:

\begin{subequations} \label{eq:eom_P+P-}
\begin{align}
\frac{\partial P_+}{\partial t} + (\mathcal{L}_1 &+ \mathcal{L}_2)(P_+ + P_-)=0 
\\
\frac{\partial P_-}{\partial t} + (\mathcal{L}_1 &+ \mathcal{L}_2)(P_+ + P_-) = 
\nonumber \\ {}  & = \Gamma_-(\vec{x};z_0)P_+ + \Gamma_+(\vec{x};z_0)P_-,
\end{align}
\end{subequations}

\noindent where $\Gamma_+(\vec{x};z_0) \equiv \Gamma_{\rm STM-NT}(\vec{x};z_0)+
\Gamma_{\rm NT-L}$, $\Gamma_-(\vec{x};z_0) \equiv \Gamma_{\rm STM-NT}(\vec{x};
z_0) - \Gamma_{\rm NT-L}$ and the Liouvillian operators $\mathcal{L}_1$ and 
$\mathcal{L}_2$ are defined as follows:

\begin{align*}
\mathcal{L}_1 & \equiv \sum_j \left[ \frac{\pi_j}{m} \frac{\partial}{\partial 
x_j} -m\omega_j^2x_j\frac{\partial}{\partial \pi_j} + \gamma_j 
\frac{\partial}{\partial \pi_j}\pi_j \right] \\
\mathcal{L}_2 & \equiv e \mathcal{E} \sum_j \varphi_j(z_0) 
\frac{\partial}{\partial \pi_j}.
\end{align*}

\noindent From Eqs. (\ref{eq:eom_P+P-}) we can derive the equations of
motion for any dynamical variable averaged over the probability densities 
$P_+$ and $P_-$: $\langle (\ldots) \rangle_\alpha \equiv \int (\ldots) 
P_\alpha (\vec{x},\vec{\pi},t) d\vec{x}d\vec{\pi}$, where $\alpha = \pm$. The 
set of equations of motion for the first moments $\langle 1 \rangle_- 
\equiv p_-(t)$, $\langle x_j \rangle_\alpha$, $\langle \pi_j \rangle_\alpha$ 
is:

\begin{subequations} \label{eq:eom}
\begin{align}
\frac{d\langle x_j \rangle_\alpha}{dt} &=\frac{\langle \pi_j \rangle_\alpha}{m}
\\
\frac{d\langle \pi_j \rangle_\alpha}{dt} & =-m\omega_j^2\langle x_j 
\rangle_\alpha -\gamma_j \langle \pi_j \rangle_\alpha + e\mathcal{E} 
\varphi_j(z_0)p_1 \\
\frac{dp_-}{dt} & = \langle \Gamma_-(\vec{x};z_0) \rangle_+ - \langle 
\Gamma_+(\vec{x};z_0) \rangle_-,
\end{align}
\end{subequations}

\noindent The set of equations (\ref{eq:eom}) is not closed because the 
exponential form of the tunneling rate $\Gamma_{\rm STM-NT}(\vec{x};z_0)$ 
introduces a coupling between the first and all the other moments. However in 
the limit of small oscillation amplitudes we can expand 
$\Gamma_{\rm STM-NT}(\vec{x};z_0)$ to first order in $x_j/\lambda$ and 
that reduces (\ref{eq:eom}) to a closed set of linear equations.

The static solution of the linearized equations of motion, $\langle \pi_j 
\rangle_\alpha = 0$, $\langle x_j^\alpha \rangle_\alpha = \bar{x}_j^\alpha$, 
$p_- = \bar{p}$, where $\bar{x}_j^\alpha$ and $\bar{p}$ are constant, describes
the nanotube as a slightly bent beam at rest. The stability of this solution 
can be investigated by substituting the expressions
$\langle \xi_k \rangle_+ \equiv \bar{\xi}_k + A_k e^{\beta_k t}$ (where $\xi_k$
is any of the dynamical variables $\langle x_k \rangle$, $\langle \pi_k 
\rangle$, $p_-$ and $A_k$ is constant) in the linearized equations of 
motion and solving for the exponents $\beta_k$ [\onlinecite{Strogatz2001}].

This procedure leads to an algebraic equation which in general cannot be 
solved analytically. However, if the dimensionless parameters 
$\varepsilon_k \equiv e\mathcal{E} \varphi^2_k(z_0)/(m \omega_k^2 \lambda)$ 
are much smaller than 1 and the nanotube is only weakly damped 
($\gamma_k \ll \omega_k$), we can look for exponents of the form 
$\beta_k \sim i\omega_k + \delta_k$, with $|\delta_k| \ll \omega_k$ and derive 
an analytical expression for the $\delta_k$ which up to the first order in  
all $\varepsilon_k$ and $\gamma_k$ reads:

\begin{equation} \label{eq:delta_k}
\delta_k = -\frac{\gamma_k}{2} + \frac{\Gamma_0^*(d)
\Gamma_{\rm NT-L}}{2\Gamma_+^*(d)}
\frac{\omega_k^2 \varepsilon_k(z_0)}{\omega_k^2 + \Gamma_+^{*2}(d)} 
\left(1 + i\frac{\Gamma_+^*(d)}{2\omega_k} \right),
\end{equation}

\noindent where $\Gamma_+^*(d) \equiv \Gamma_0^*(d)+\Gamma_{\rm NT-L}$. 
The condition $\varepsilon_k \ll 1$ can be taken as a definition of the weak 
electromechanical regime, since it implies that the shift in the
equilibrium position of the nanotube at point $z_0$ when it is 
charged by one extra electron, 
$\delta u \equiv e\mathcal{E}\sum_j \varphi_j^2(z_0)/(m \omega_j^2)$, is 
much smaller than the tunneling length $\lambda$. 
For realistic values of the parameters which are consistent with 
the conditions of validity of our model ($V \sim$ 0.1 V, the 
capacitances of the tunnel junctions 
$C_{\rm STM-NT} \sim C_{\rm NT-L} \sim 10^{-18}$ F as reported in 
[\onlinecite{LeRoy2004a}], 
$\lambda \sim 10^{-10}$ m, $\omega_k \sim 10^2$ MHz), $m \sim 10^{-22}$ kg, $L \sim 10^{-7}$ m,  
the regime of weak electromechanical coupling is attained: 
$\varepsilon_k(z_0) \lesssim 0.1$ for all the modes at every position $z_0$.
%the electromechanical coupling results actually weak: $\delta u / \lambda \simeq 0.1$.

The sign of the real part of $\delta_k$ in Eq. (\ref{eq:delta_k}) determines 
the stability of the static solution for the $k-$th average mode amplitude. 
If $\mathfrak{Re}[\delta_k]>0$ then  $\langle x_k \rangle_+$ increases 
exponentially in time, hence the static solution for the $k$-th mode is 
unstable. This is the signature of a ``shuttle-like'' electromechanical 
instability. On the other hand, if $\mathfrak{Re}[\delta_k]<0$ the energy 
pumped into the vibrational mode by the electrostatic field is not able to 
compensate the loss due to dissipation and after a time interval of the order 
of $1/\gamma_k$ the $k$-th mode amplitude decays to its static value. 

For fixed values of $\gamma_k$, $\mathcal{E}$ and $\Gamma_2$, the sign of 
$\mathfrak{Re}[\delta_k]$ becomes a function of $z_0$ and $\Gamma_0^*(d)$, i.e.
it depends only on the position of the STM tip in the $xz$ plane. The set of 
values of $z_0$ and $\Gamma_0^*(d)$ for which the real part of $\delta_k$ is 
positive defines the instability region for the $k$-th mode amplitude in the 
plane ($z_0$,$\Gamma_0^*(d)$). 

In order to map the instability regions we have to specify an analytic expression 
for the damping rates $\gamma_k$. The dissipation of mechanical energy in NEMS 
can take place through several mechanisms. However, in spite of the variety of
the dissipative processes in solids, their effect on the NEMS performance can
be described by considering the retardation induced in the NEMS response to 
mechanical perturbations (which adds to the ``instantaneous'' elastic behaviour). 

In order to include this effect in our model we follow the approach introduced by Zener 
and formally replace the Young modulus with a frequency dependent complex 
function which results in the following expression for the damping rates: 
$\gamma_k \equiv Y \omega_k^2\tau /(1+\omega_k^2 \tau^2)$ [\onlinecite{Cleland2003}]. 

A large class of dissipative phenomena in solids (e.g. thermoelasticity, 
dislocations and defects dynamics) can be parametrized though the 
dimensionless coefficient $Y$ and the relaxation time $\tau$, which both depend 
on temperature as well as on several geometric and material properties
[\onlinecite{Nowick1972}]. 

\begin{figure} 
\includegraphics[width=0.5\textwidth]{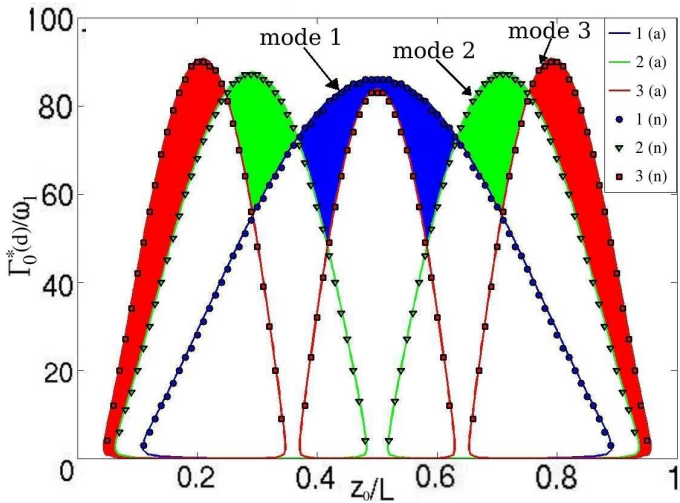} 
\caption{(Color online) Regions of instability in the parameter plane ($z_0$, 
$\Gamma_0^*(d)$) for the first three bending modes when the 
inverse characteristic time of the nanotube mechanical response to the external 
force is smaller than the frequencies of the nanotube vibrational modes,
$1/\tau = 0.1\omega_1$. The regions computed from the real part of the analytic 
expression (\ref{eq:delta_k}) (solid lines) are compared with the results 
obtained by the numerical solution of the linearized equations of motion 
(markers). The parts of the plot filled with colors correspond to the 
values of ($z_0$, $\Gamma_0^*(d)$) for which a single vibrational mode is 
excited. The other relevant parameters are $e\mathcal{E}/(m\lambda \omega_1^2)=0.1$, 
$\Gamma_{\rm NT-L}=5 \omega_1$, $Y =10^{-4}$.}
\label{fig:big_omegatau}
\end{figure}

We first consider the limit in which the characteristic inverse time 
of the retarded mechanical response is much smaller than the 
frequencies of the nanotube eigenmodes: $1 /\tau \ll \omega_k$. 
In this case the damping term in Eq. (\ref{eq:delta_k}) 
does not depend on the frequency and the dissipation rate 
is the same for all the modes. 

In Fig. (\ref{fig:big_omegatau}) the instability regions determined by Eq. 
(\ref{eq:delta_k}) for the first three modes are plotted together with the 
real parts of the exponents $\beta_k$ obtained from the numerical analysis of 
the linear stability problem. The areas filled with colours correspond to the 
values of ($z_0$, $\Gamma_0^*(d)$) for which only a single vibrational mode is 
unstable, while the regions where two or three modes are unstable are left 
blank. 

\begin{figure}
\includegraphics[width=0.5\textwidth]{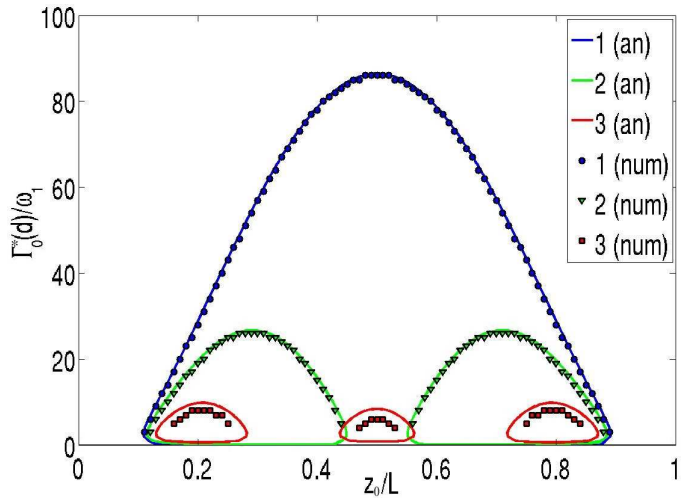} 
\caption{(Color online) Regions of instability in the parameter plane ($z_0$, 
$\Gamma_0^*(d)$) for the first three bending modes when the inverse characteristic 
time of the nanotube mechanical response to the external force is 
larger than the frequencies of the nanotube vibrational modes,
$1/\tau = 10\omega_1$. The regions computed from the real part of the analytic 
expression (\ref{eq:delta_k}) (solid lines) are compared with the results 
obtained by the numerical solution of the linearized equations of motion
(markers). The region of instability for the first mode includes all the 
values of ($z_0$, $\Gamma_0^*(d)$) for which the other modes 
are unstable, therefore no selective excitation can be attained in this regime.
The other relevant parameters are $e\mathcal{E}/(m\lambda \omega_1^2)=0.1$, 
$\Gamma_{\rm NT-L}=5 \omega_1$, $Y =10^{-4}$.}
\label{fig:small_omegatau}
\end{figure}

The physical picture presented in Fig. (\ref{fig:big_omegatau}) changes 
drastically in the opposite limit, $1/\tau \gg \omega_k$, as can be seen in Fig. 
(\ref{fig:small_omegatau}). In this case the first mode is characterized
by the smallest dissipation rate, $\gamma_1 \ll \gamma_k \quad k \neq 1$, 
therefore if any of the modes is unstable, also the first one is unstable.
That excludes the possibility of promoting a selective instability in the 
limit $1/\tau \gg \omega_k$.

The dynamical behaviour of the nanotube in the regime of single-mode instability is 
qualitatively the same of the ordinary ``shuttle'' system [\onlinecite{Gorelik1998}]. 
The amplitude of the oscillations increases exponentially until it reaches a 
certain steady value which depends on the parameters of the system. This transition 
is characterized by a large enhancement of the current (with respect to the static 
tunneling regime) that can be experimentally detected by measuring the current 
flowing through the device for different positions of the STM tip. 

In the phase space of the system the dynamical state in this situation is described 
by a limit cycle, that is an isolated closed trajectory characterized by finite amplitude 
oscillations [\onlinecite{Strogatz2001}]. Here we have shown that the frequency  
of this stable self-oscillating state can be selected among the whole set of 
nanotube resonant frequencies through an accurate positioning of the STM tip. 

In conclusion in the present work we studied the dynamics of the flexural 
vibrations of a suspended carbon nanotube in which extra electrons are injected
at a position-dependent rate. We showed that a localized constant electrostatic
field can excite many transverse vibrational modes of the nanotube into a 
``shuttle-like'' regime of charge transport. For a fixed bias voltage and in 
presence of dissipative processes with inverse characteristic times much smaller
than the frequencies of the nanotube vibrational modes, we found that it is 
possible to induce a selective instability through an accurate positioning of a STM. 
It thus seems possible to extend the approach followed here to other systems 
characterized by a non trivial coupling between charge transport and mechanical 
degrees of freedom.

The author wants to thank L. Y. Gorelik, R. I. Shekhter and M. Jonson for 
fruitful discussions and support. Partial financial support from the Swedish VR
and from the Faculty of Science at the University of Gothenburg through its 
``Nanoparticle'' Research Platform is gratefully acknowledged.

\bibliography{/home/santandr/Papers/Paper2/Biblio}

\end{document}